\documentclass{ifacconf}

\usepackage{color,xcolor}
\usepackage{graphicx}
\usepackage{natbib}

\newcommand{\st}{\mathrm{subject\ to}}
\newcommand{\argmin}{\mathop{\mathrm{argmin}}}
\newcommand{\argmax}{\mathop{\mathrm{argmax}}}
\newcommand{\maximize}{\mathop{\mathrm{maximize}}}
\usepackage{algorithm}
\usepackage{textcomp}
\usepackage{savesym}
\savesymbol{AND}
\usepackage{algorithmic}

\usepackage{amssymb}
\usepackage{subfigure}
\usepackage{amsmath}
\usepackage{url}
\usepackage{multirow}
\usepackage{braket}
\newcommand{\tabincell}[2]{\begin{tabular}{@{}#1@{}}#2\end{tabular}}

\newcommand{\BQIC}{Berkeley Center for Quantum Information and Computation, Berkeley, California 94720 USA}
\newcommand{\DeptMath}{Department of Mathematics, University of California, Berkeley, California 94720 USA}

\newcommand{\LBLMath}{Computational Research Division, Lawrence Berkeley National Laboratory, Berkeley, CA 94720, USA}
\newcommand{\DeptChem}{Department of Chemistry, University of California, Berkeley, California 94720 USA}
\newcommand{\SC}{SC Solutions, Sunnyvale, California 94085 USA}

\begin{document}

\begin{frontmatter}

\title{Robust Control Optimization for Quantum  Approximate Optimization Algorithm} 

\author{Yulong Dong$^{1,2}$} 
\author{Xiang Meng$^{3}$}
\author{Lin Lin$^{3,4}$}
\author{Robert Kosut$^5$}
\author{K. Birgitta Whaley$^{1,2}$}

$^1$\BQIC\ 
$^2$\DeptChem\ 
$^3$\DeptMath\ 
$^4$\LBLMath\ 
$^5$\SC

\begin{abstract}

Quantum variational algorithms have garnered significant interest recently, due to their feasibility of being implemented and tested on noisy intermediate scale quantum (NISQ)
devices. We examine the robustness of the quantum approximate optimization
algorithm (QAOA), which can be used to solve certain quantum control problems, state preparation problems, and combinatorial
optimization problems. 
We demonstrate that the error of QAOA simulation can be significantly reduced by robust control optimization techniques, specifically, by sequential convex programming (SCP), to ensure error suppression in situations where the source of the error is known but not necessarily its magnitude. 
We show that robust
optimization improves both the objective landscape of QAOA as well as
overall circuit fidelity in the presence of coherent errors and errors in initial state preparation. 
\end{abstract}


\end{frontmatter}

\section{Introduction}
Achievements in quantum information and computing have boosted prospects towards solving complicated problems that challenge classical computational resources. However, the performance of near term quantum devices can be severely limited by decoherence effects. To leverage the computational power of near term quantum devices while also operating within their constraints, a number of hybrid quantum-classical algorithms have been proposed in which a quantum component is embedded within a classical processor.  
In particular, quantum variational algorithms (QVA), which embed a quantum protocol dependent on some tunable parameters inside a classical optimizer, have received significant amount of attention in the past few years. 
Examples of QVA include the variational quantum eigensolver(\cite{mcclean2016theory}), the quantum approximate optimization algorithm (QAOA)(\cite{farhi2014quantum}), and quantum variational autoencoders (\cite{KhoshamanVinciDenisEtAl2018}). The basic idea of QVA is to parameterize the quantum wavefunctions of interest using a relatively small set of parameters and to use a quantum state circuit to prepare the state. For each given parameter set, corresponding to a parameterized set of gates in the circuit, the objective function can be efficiently evaluated on a quantum computer through intrinsically noisy quantum measurements. The parameters are then variationally adjusted  and optimized on a classical computer, according to the outcomes of the quantum measurements.

In this paper, we focus on the quantum approximate optimization
algorithm (QAOA), which is one such hybrid quantum-classical algorithm that provides a compact variational ansatz for a quantum state evolved under a time dependent Hamiltonian by acting on a initial state by switching between two specifically designed control Hamiltonians. The universality (\cite{Lloyd1995,lloyd2018quantum}) and optimality of QAOA for certain problems (\cite{YangRahmaniShabaniEtAl2017}) have led to many applications of this approach in quantum information processing, ranging from a broad array of discrete combinatorial problems (\cite{farhi2014quantum,lloyd2018quantum,1805.03265}), to more recent extensions addressing continuous variable problems (\cite{1902.00409}).

Finding efficient classical algorithms for the optimization of parameters in QAOA is currently an active area of research.  Methods under investigation include approaches using gradient optimization (\cite{ZhuRabitz1998,MadayTurinici2003}),  Pontryagin's maximum principle  (\cite{YangRahmaniShabaniEtAl2017,BaoKleerWangEtAl2018}), and reinforcement learning (\cite{bukov2018reinforcement,NiuBoixoSmelyanskiyEtAl2019}). Due to the variational nature of the algorithm, first order changes of parameters  may only lead to second order changes of the fidelity. In this sense, all QVAs are to some extent naturally resilient to the impact of errors and noise. 
The source of error may originate from coupling to the environment, uncertain duration of the propagation time, and measurement errors. 
In this work, we show that even if only the source of the error and its range of magnitude are known, \textit{i.e.}, but not the exact magnitude, one can nevertheless significantly improve the fidelity of QAOA based simulation. 
We employ methods of robust optimal control, and in particular the sequential convex programming (SCP) approach (\cite{EE364b}) to improve the overall circuit robustness against errors. In this work we focus on coherent errors and initial state errors. Future work will address mitigation against environmental errors.

Section \ref{sect:qaoa} reviews preliminary concepts and notations regarding QAOA and the uncertainty in the Hamiltonian. We formulate the robust optimization algorithm for QAOA and present an SCP approach to solve this in Section \ref{sect:scp}. In Section \ref{sect:errmitigate} we present several numerical applications of QAOA with the resulting robust control, including both single-qubit and multi-qubit examples. Section \ref{sect:sum} concludes and presents an outlook for further work.

\section{QAOA and Hamiltonian Uncertainty}\label{sect:qaoa}
By iteratively applying two Hamiltonian operators, QAOA provides control schemes in a hybrid quantum-classical manner, which can
be implemented on near term quantum devices. Starting from an initial state $\ket{\psi_i}$, QAOA prepares a state $\ket{\psi_{\theta}}$ using the following ansatz
\begin{equation} \label{eq:qaoa}
   \ket{\psi_{\theta}} = U(\theta) \ket{\psi_i} := \prod_{j = 1}^p U(H_B, \theta_j^B) U(H_A, \theta_j^A) \ket{\psi_i}.
\end{equation}
Here $U(H, \theta) = e^{- i H \theta}$. 
The evolution operator $U({\theta})$ is therefore the product of unitary transformations generated by two Hamiltonian operators alternatively, and $p$ is called the {depth} of QAOA.

QAOA can be rewritten compactly in terms of a time-dependent Schr\"odinger equation, with the Hamiltonian $H(t) = s(t) H_A + [1 - s(t)] H_B$. The control parameter $s(t)$ takes only the values $0$ or $1$, thus corresponding to a bang-bang control. The optimization variables are the sequence of time durations $\theta = (\theta_1^A, \theta_1^B, \cdots, \theta_p^A, \theta_p^B)$. A commonly used objective function is the fidelity, defined as the overlap between $\ket{\psi_{\theta}}$ and a target state $\ket{\psi_t}$, \textit{i.e.},
\begin{equation}
        \mathcal{F}(\theta) = | \langle \psi_i | U(\theta) | \psi_t \rangle |^2.
\end{equation}
 Previous research suggests that the control landscape of QAOA can be highly complex (\cite{StreifLeib2019,NiuLuChuang2019}). Although the need for error mitigation has been recognized, most studies of QAOA to date, however, neglect effects of noise that undermine the robustness of QAOA control. In this work, we consider a robust QAOA scheme to improve overall circuit fidelity against uncertainty originating from the Hamiltonian. More specifically, we assume the Hamiltonian takes the form $H(\delta) = \bar{H}+\tilde{H} (\delta)$, where $\bar{H}$ is the original Hamiltonian operator in the absence of noise, and $\tilde{H}$ is due to the noise. The undetermined value of parameter
$\delta$ indicates the magnitude of the noise, with the value of $\delta$ only known to within a prescribed range $\Delta$. The goal of robust QAOA optimization is to find a sequence of angles (pulse times) that achieves high fidelity for any realization in the uncertainty set $\Delta$, equivalently a robust bang-bang control.

\section{Sequential Convex Programming}\label{sect:scp}
The robust QAOA optimization problem can be formulated as the following max-min problem:
\begin{equation}
\label{eq:control}
\begin{array}{ll}
\maximize_\theta \quad & \min_\delta \ \mathcal{F}(\theta, \delta),\\
\st \quad & \theta \in \Theta, \quad \delta \in \Delta.
\end{array}
\end{equation}
Here
\begin{equation}
        \mathcal{F}(\theta, \delta) = | \langle \psi_i | U(\theta, \delta) | \psi_t \rangle |^2,
\end{equation}
 $\theta$ stands for the control protocol within a convex feasible set $\Theta$, and $\delta$ represents the uncertainty. Eq.\eqref{eq:control} thus maximizes the worst case fidelity within $\Delta$. We do not require  the set of uncertainty parameters $\Delta$ to be convex. 

Various methods have been proposed to solve the robust optimization problem. For instance, b-GRAPE and a-GRAPE methods (\cite{wu2019learning,ge2019robust}) are variants of the well-known GRAPE method (\cite{khaneja2005optimal}), which was designed for quantum control in the absence of uncertainty. In order to enhance the robustness of the solutions, b-GRAPE obtains the update of $\theta$ with respect to randomly generated samples $\delta\in \Delta$, while a-GRAPE performs optimization with respect to samples $\delta\in\Delta$ generated from an adversarial procedure. In this paper, we utilize sequential convex programming (SCP) (\cite{kosut2013robust,allen2017optimal}) to solve the max-min problem in Eq.\eqref{eq:control} and compare the performance with that of the a- and b-GRAPE variants.

SCP iteratively solves a subproblem, where the objective function is locally approximated to be a concave function. For instance, we may approximate $\mathcal{F}$ by a linear function
\begin{equation}\label{scp:gradapp}
\mathcal{F}(\theta + \tilde{\theta}, \delta) \approx \tilde{\mathcal{F}}(\tilde{\theta}; \theta, \delta)= \mathcal{F}(\theta, \delta) + \tilde{\theta}^T \nabla_\theta \mathcal{F}(\theta, \delta).
\end{equation}
Alternatively, the information of Hessian can be utilized to obtain a more accurate approximation
\begin{equation}\label{scp:hessapp}
\tilde{\mathcal{F}}(\tilde{\theta}; \theta, \delta) = \mathcal{F}(\theta, \delta) + \tilde{\theta}^T \nabla_\theta \mathcal{F}(\theta, \delta) + \frac{1}{2} \tilde{\theta}^T \mathrm{Hess}_-(\theta, \delta) \tilde{\theta},
\end{equation}
where $\mathrm{Hess}_-$ represents the negative semidefinite part of the Hessian of the fidelity, so that $\tilde{\mathcal{F}}(\tilde{\theta}; \theta, \delta)$ is a concave function. To carry out an approximate optimization on the original problem, a convex trust region $\Theta_{\mathrm{trust}}$ is introduced that guarantees the accuracy of the approximation. The size of the trust region is flexible and 
updated on-the-fly via the trust region algorithm (\cite{EE364b}).  The minimization over uncertainty is implemented by sampling realizations $\{ \delta_i : i = 1, \cdots, L \}$ from the uncertainty set $\Delta$ to provide a numerical resolution of this. A summary of the SCP approach is provided in Algorithm.\ref{alg:scp}.
\begin{algorithm}[!htbp]
        \caption{Robust optimization through sequential convex programming}
        \label{alg:scp}
        \begin{algorithmic}
                \REQUIRE initial guess $\theta_0 \in \Theta$, initial trust region $\Theta_{\mathrm{trust}}$, realizations $\{ \delta_i \in \Delta : i = 1, \cdots, L \}$; parameter $0<\eta_2<\eta_1<1$, $0<\gamma_2<1<\gamma_1$, a positive integer $t_{max}$, stop criteria $\text{tol}_{d}$ and $\text{tol}_{\sigma}$.
                \ENSURE optimal control vector $\theta$.
                \FOR{$t=1,2,\dots,t_{max}$}
                        \STATE (1) Compute fidelities, gradients and Hessians associated with each uncertainty realization.
                        \STATE (2) Approximate the fidelity $\mathcal{F}(\theta,\delta)$ by either Eq.\eqref{scp:gradapp} or Eq.\eqref{scp:hessapp}.  Solve increment $\tilde{\theta}$ from the convex problem
                        \begin{equation}
                        \begin{array}{ll}
                        \maximize \quad & \displaystyle \min_{i = 1, \cdots, L} \tilde{\mathcal{F}}(\tilde{\theta}; \theta, \delta_i),\\
                        \st \quad & \theta + \tilde{\theta} \in \Theta, \quad \tilde{\theta} \in \Theta_{\mathrm{trust}}.
                        \end{array}
                        \end{equation}
                        \STATE (3) Calculate the ratio
                        \begin{equation}
                            \sigma=\frac{\min_i \mathcal{F}(\theta+\tilde \theta,\delta_i)-\min_i \mathcal{F}(\theta,\delta_i)}{\min_i \tilde{\mathcal{F}}(\tilde{\theta}; \theta, \delta_i)-\min_i\mathcal{F}(\theta,\delta_i)}.
                        \end{equation}
                        \IF{$\sigma>0$}
                        \STATE{Update $\theta=\theta+\tilde\theta$.}
                        \ENDIF
                        
                        Update $d$, the diameter of trust region $\Theta_{\mathrm{trust}}$ by following rule:
                            \begin{equation}
                                d=\left\{\begin{array}{rl}
                                    \gamma_1d & \qquad\text{if } \eta_1<\sigma, \\
                                    d &\qquad \text{if } \eta_2\le \sigma\le \eta_1, \\
                                    \gamma_2d & \qquad\text{if } \sigma<\eta_2.
                                \end{array}\right.
                            \end{equation}
                        \IF{$d\in(0,\text{tol}_d)$ or $\sigma\in(0,\text{tol}_{\sigma})$} \STATE \textbf{break} 
                        \ENDIF
                \ENDFOR. 

        \end{algorithmic}
\end{algorithm}

After introducing a slack variable, $f_0$, the convex problem proposed in Algorithm \ref{alg:scp} can be rewritten in a standard form as\begin{equation}
\label{eq:std-convex}
\begin{array}{ll}
\maximize \quad & f_0,\\
\st \quad &\tilde{\mathcal{F}}(\tilde{\theta}; \theta, \delta_i) \ge f_0,\quad \forall i = 1, \cdots, L\\
& \theta + \tilde{\theta} \in \Theta, \quad \tilde{\theta} \in \Theta_{\mathrm{trust}}.
\end{array}
\end{equation}
If $\tilde{\mathcal{F}}(\tilde{\theta}; \theta, \delta)$ is a linear approximation, and $\Theta$, $\Theta_{\mathrm{trust}}$ are box-constrained sets, Eq.\eqref{eq:std-convex} is a linear programming problem. In addition, if $\Theta$ and $\Theta_{\mathrm{trust}}$ are either box- or ball-constrained sets, then regardless of the approximation chosen for $\tilde{\mathcal{F}}(\tilde{\theta}; \theta, \delta)$, Eq.\eqref{eq:std-convex} can always be converted to a quadratic constrained programming. In both cases, a standard optimizer such as Gurobi (\cite{gurobi}) can be utilized to solve the subproblem very efficiently.

\section{Robust optimization for error mitigation}\label{sect:errmitigate}
\subsection{Single-qubit system\label{sect:numeric:sing}}
As a first example, we investigate the use of robust QAOA to transfer the state of a single qubit under Hamiltonian uncertainty. Specifically, given an initial qubit state $|\psi_i\rangle$ that is the ground state of $H_i = - \sigma^z + 2 \sigma^x$, we adopt a QAOA Hamiltonian ansatz with the following two Hamiltonians to transfer the system state to $|\psi_t\rangle$, the ground state of $H_t = - \sigma^z - 2 \sigma^x$:
\begin{equation}
        H_A(\omega_A) = - \sigma^z + \omega_A \sigma^x, \quad H_B(\omega_B) = - \sigma^z + \omega_B \sigma^x
\end{equation}
Here the uncertainty derives from uncertainties in the parameters $\omega_A, \omega_B$, \textit{i.e.}, $\delta = (\omega_A, \omega_B) \in \Delta$. The reference control Hamiltonians without uncertainty are assumed to be those with $(\omega_A, \omega_B) = (4, -4)$. We shall study the optimal solution obtained for various sizes of the uncertainty set $\Delta$.

\begin{figure}
        \centering
        \subfigure[\label{fig:sing-fid-p}]{
                \includegraphics[width=8cm]{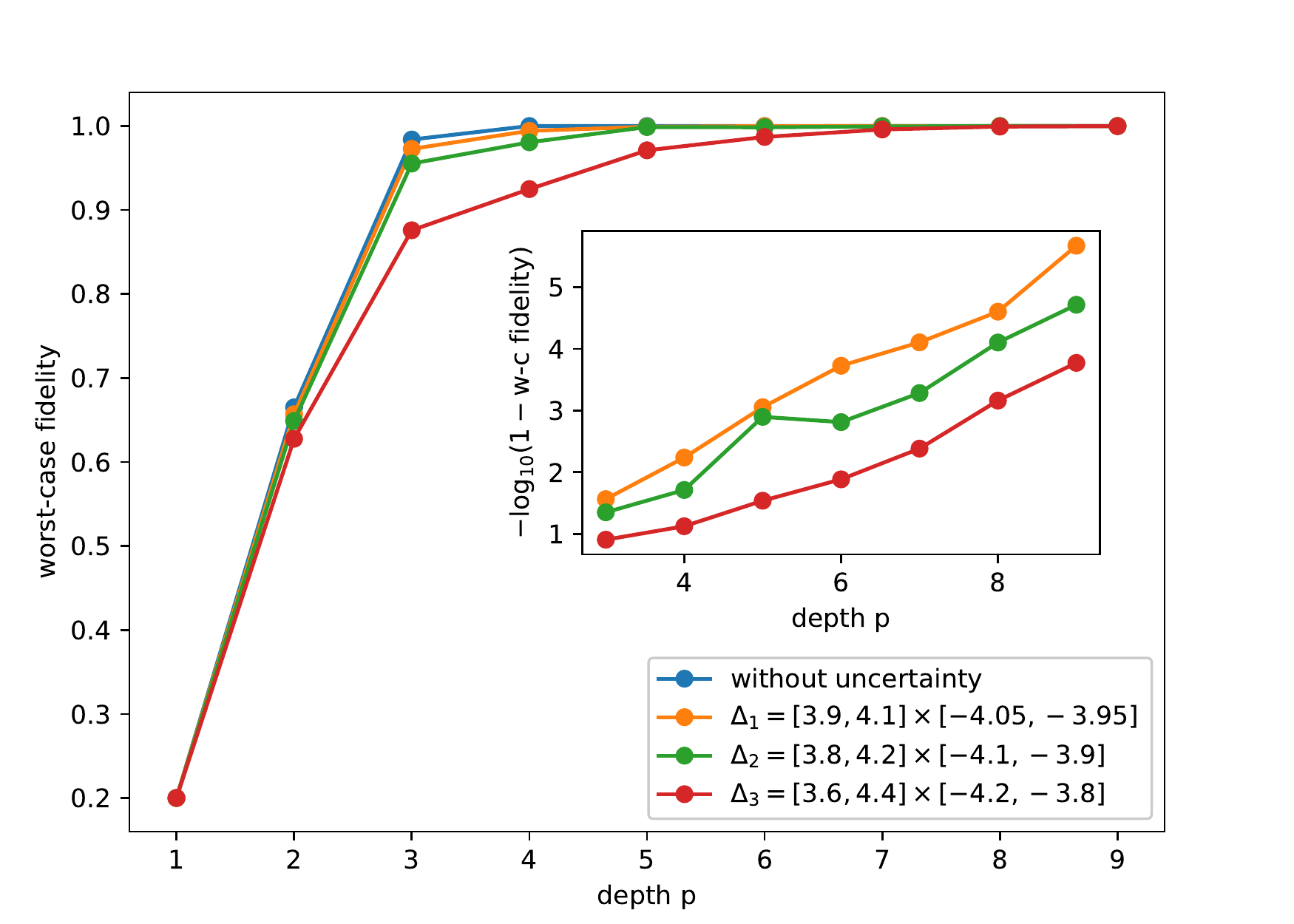}
        }
        \subfigure[\label{fig:sing-fidcontour}]{
                \includegraphics[width=10cm]{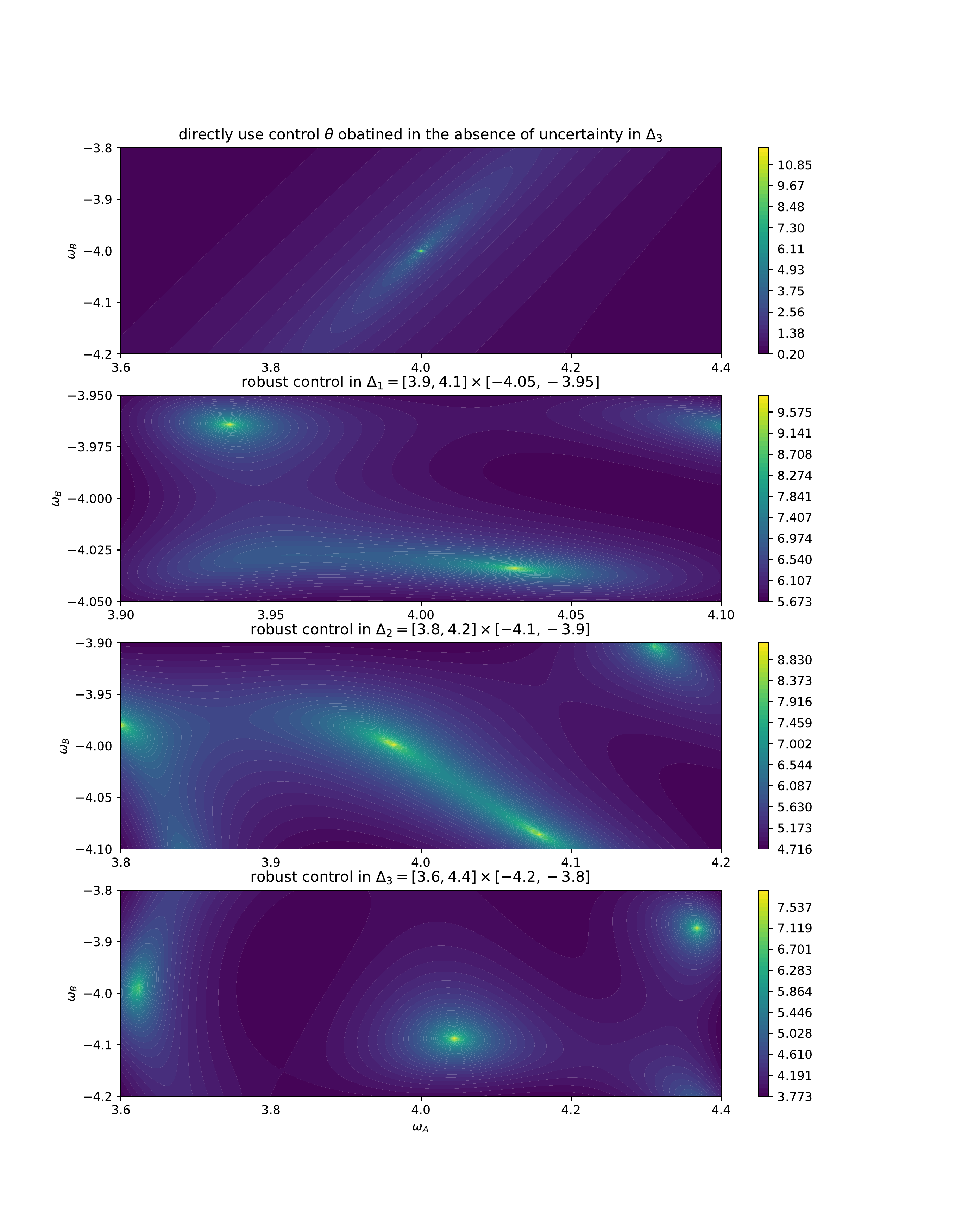}
        }
        \caption{Robust QAOA for the single-qubit system. (a) Optimal worst-case fidelity as the function of QAOA depth. Different curves correspond to different uncertainty models, $\Delta_1 \subset \Delta_2 \subset \Delta_3$. (b) The landscape of the infidelity $- \log_{10}(1 - \mathcal{F})$.\label{fig:sing}}
\end{figure}

The fidelity for state transfer obtained by robust QAOA in this single-qubit control problem is displayed in Fig.\ref{fig:sing} for several values of $\Delta$ and for the uncertainty-free Hamiltonian (blue dots). Panel a) shows that the fidelity increases as the depth, \textit{i.e.}, number of layers $p$ in the QAOA iteration, increases. This agrees with the fact that a solution associated with greater depth includes that of lower depth, with the extra entries set to zero. Furthermore, since the uncertainty sets are increasing in size, namely $\Delta_1 \subset \Delta_2 \subset \Delta_3$, the numerical results in Fig.\ref{fig:sing-fid-p} also demonstrate that the robustness of QAOA decreases as the system becomes less certain, leading to a lower achievable fidelity. Starting from optimal QAOA sequences without uncertainty, contour maps of fidelity on the uncertainty set are shown in Fig.\ref{fig:sing-fidcontour}. These exhibit multi-peak patterns not centered at the uncertainty-free point $(4, -4)$. This confirms that the robust optimization modifies the control without uncertainty to values that are optimal for the worst-case fidelity (max-min). The distinct terrain structures evident in Fig.\ref{fig:sing-fidcontour} also show that different optimal control solutions are obtained with different ranges of uncertainty $\Delta$. This single-qubit example shows that robust optimization of a QAOA control protocol with sufficient depth enables high fidelity of quantum state transfer to be consistently attained for all uncertainty realizations. The multi-peak patterns also give an indication of how different systems designs can affect robustness. 

\subsection{Multi-qubit system (I)}\label{sect:numeric:manipulation}
We now consider the robust QAOA for a multi-qubit system with the following Hamiltonian,
\begin{equation}
\begin{split}
    H(h; \omega_1, \omega_2) =& - \left( 1 + \omega_1 \right) \sigma_1^z \sigma_2^z - \left( 1 + \omega_2 \right) \sigma_2^z \sigma_3^z\\
    &- \sum_{j = 3}^{N-1} \sigma_j^z \sigma_{j + 1}^z - \sum_{j = 1}^N \left[\sigma_j^z + h \sigma_j^x \right].
\end{split}
\end{equation}
Here the uncertainty parameters $\omega_{1}, \omega_2$ affect the couplings of the first two pairs of qubits. The spin operator $\sigma^z$ is coupled to a constant external field to avoid the antiferromagnetic-paramagnetic phase transition (\cite{bukov2018reinforcement}). The configuration of the chain is controlled by the tunable magnetic field $h$. Previous work has investigated the optimal control problem in the absence of uncertainty using reinforcement learning (RL) (\cite{bukov2018reinforcement}). In that approach a RL agent makes a decision on a boolean action space $h \in \{ -4, +4 \}$ within each time interval. To improve the robustness of QAOA with control Hamiltonians taking $h = -4\ \mathrm{or}\ +4$ against uncertainty in the couplings $\omega_1$ and $\omega_2$, we employ the robust optimization algorithm to further adjust the control angles over uncertainty samples. The initial state $\ket{\psi_i}$and target state $\ket{\psi_t}$ are set to be the ground states at field values $h = -2\ \mathrm{and}\ +2$ respectively in the absence of uncertainty.

\begin{figure}
    \centering
    \includegraphics[width=9.5cm]{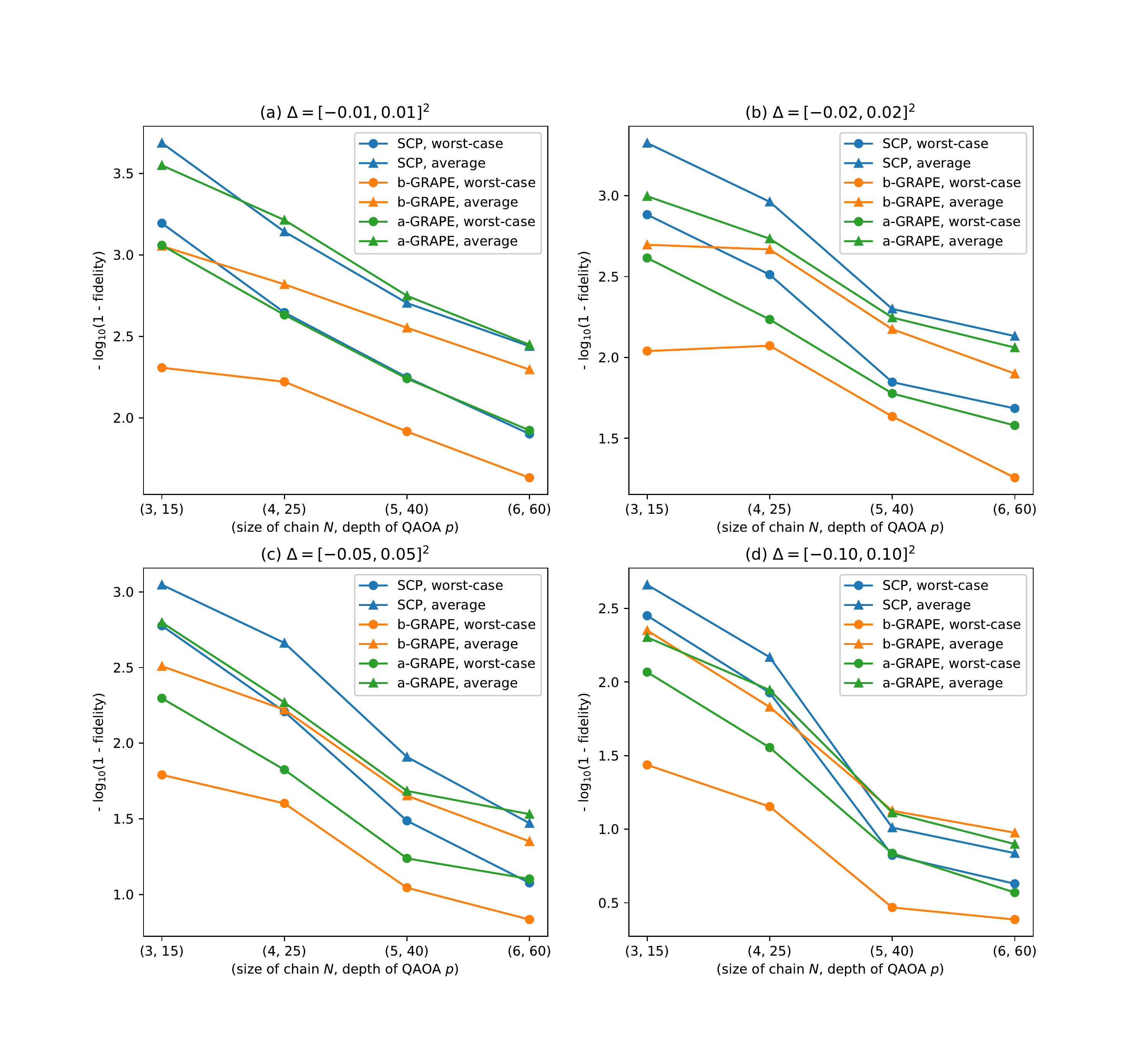}
    \caption{Robust QAOA for multi-qubit system (I) for different system sizes, QAOA depths and uncertainty sets.}
    \label{fig:multi-manipulation}
\end{figure}

Results of robust control for this multi-qubit system using SCP, b-GRAPE and a-GRAPE are shown in Fig.\ref{fig:multi-manipulation}.
The objective functions in a-GRAPE and SCP calculations are the worst-case fidelity, while the objective function of b-GRAPE is the average fidelity. The optimized controls in each case are then used to evaluate the fidelity of the other kind. We observe that the fidelity generally decreases as the system size, \textit{i.e.}, the number of qubits $N$ in the chain, is increased, despite a simultaneous increase of the QAOA depth $p$. Comparison of the relative performance of the three different optimization algorithms shows that in nearly all cases, the worst-case fidelity obtained using SCP surpasses that achieved using b-GRAPE or a-GRAPE. Furthermore, in the situations where one of the GRAPE algorithms gives similar fidelities as SCP, the SCP algorithm is found to be more efficient, as illustrated by the timing data shown in Tab.\ref{appendix:compare}.\ref{tab:manipulation}.

\subsection{Multi-qubit system (II)}\label{sect:numeric:multi}
Here we consider a second state transfer problem for a multi-qubit system. Similar to binary sequences in a classical computer, the model is defined on a one-dimensional chain consisting of $N$ qubits (\cite{NiuLuChuang2019}). The notation $|\bar{k}\rangle = |0\rangle_1 \cdots |1\rangle_k \cdots |0\rangle_N$ means that the $k^{th}$ site is excited with respect to the $z$-basis. The goal of the control is to move an excitation all the way along the chain, namely, $|\bar{1}\rangle \rightarrow |\bar{N}\rangle$. The following two Hamiltonians are utilized as the ansatz for QAOA iteration:
\begin{equation}
        \begin{aligned}
                H_A(\delta) &= \sum_{i = 1}^{N - 1} (\sigma^x_i \sigma^x_{i + 1} + \sigma^y_i \sigma^y_{i + 1}) + \delta \sigma^z_{\lfloor \frac{N}{2} \rfloor - 1} \sigma^x_{\lfloor \frac{N}{2} \rfloor} \sigma^z_{\lfloor \frac{N}{2} \rfloor + 1}\\
                H_B &= \frac{1}{2} (\sigma^z_N + I_N).
        \end{aligned}
\end{equation}
Here $H_B$ projects onto the target state, and $H_A(0)$, the off-diagonal Hamiltonian without uncertainty, introduces a swap operation between neighboring qubits. The three-qubit coupling term whose strength can take values within a bounded set $\delta \in \Delta$ provides a Hamiltonian uncertainty and the time evolution operator is thus dependent on this uncertainty.

As in the previous problems, for each instance of $\delta$, the fidelity is defined using the overlap between target state and the state generated by QAOA iteration,
\begin{equation}
        \mathcal{F}(\theta, \delta) = |\langle\bar{N}| U(\theta, \delta) |\bar{1}\rangle|^2.
\end{equation}

Since the length of the chain $N$ varies, we shall always consider QAOA with depth $p = N + 1$. We set the parameter set of Hamiltonian uncertainty to be $\Delta = [- 0.15, 0.15]$.

\begin{figure}
        \centering
        \includegraphics[width = 9.5 cm]{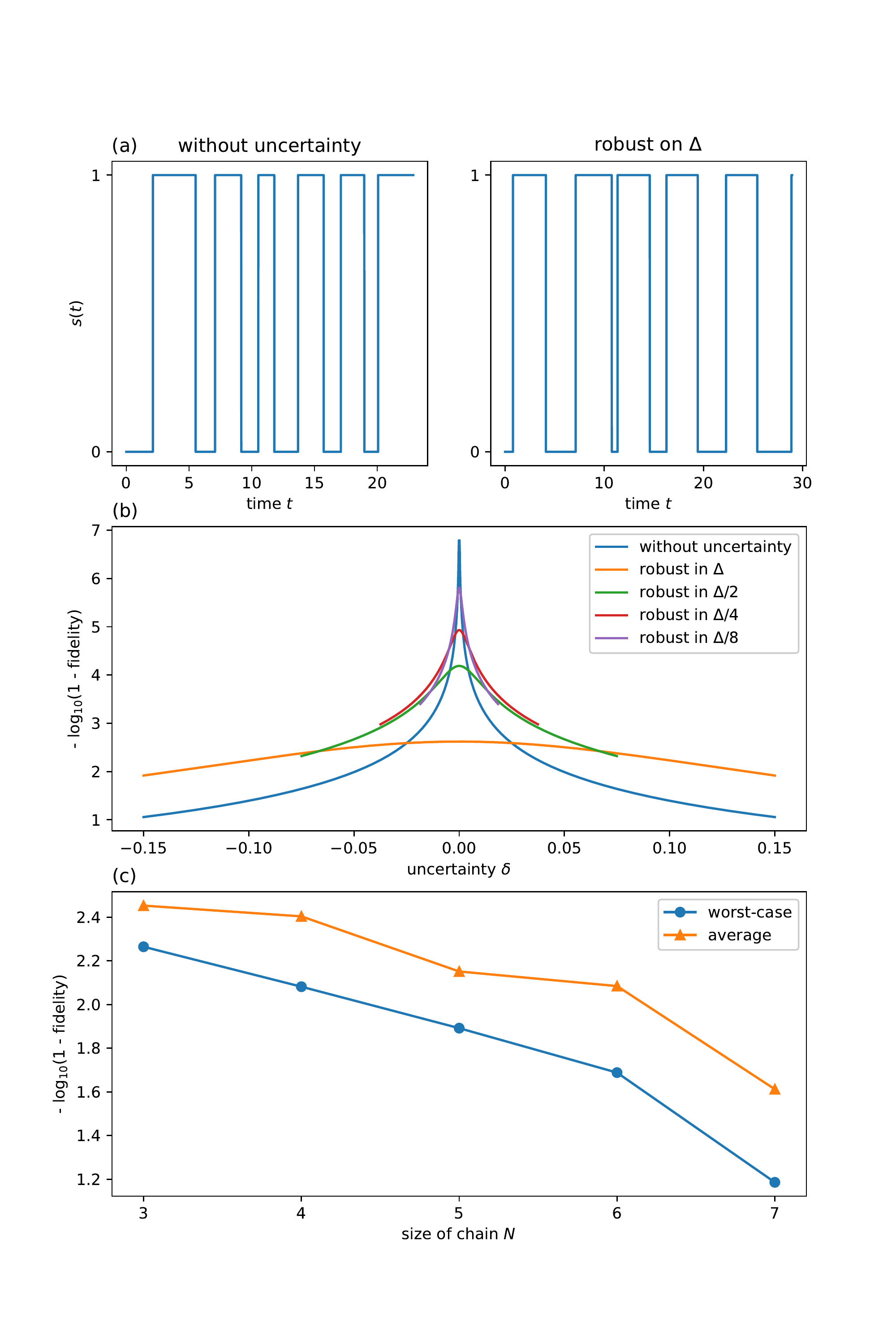}
        \caption{Robust QAOA for multi-qubit system (II). (a) Comparison of control protocols of five-qubit chain and six-depth QAOA. Left: optimal bang-bang protocol solving optimization problem without uncertainty, which is used as initial guess in SCP. Right: optimal bang-bang control obtained through SCP with samples from $\Delta$. (b) Optimal fidelity of five-qubit chain and six-depth QAOA as a function of uncertainty set $\Delta = [-0.15, 0.15], \Delta/2, \Delta/4, \Delta/8$. (c) Optimal fidelities of qubit chains with different sizes but fixing $\Delta = [-0.15, 0.15]$.\label{fig:multi}}
\end{figure}

Fig. \ref{fig:multi}(a) shows a comparison between the optimal control variables in the absence of uncertainty $\theta^{\mathrm{w.o.}}$ and the corresponding robust optimal control variables that are generated by a subsequent robust optimization from the initial guess $\theta^{\mathrm{w.o.}}$. This reveals significantly different structures, and thus implies that a distinct control protocol is produced after introducing uncertainty. Fig. \ref{fig:multi}(b) shows that when the uncertainty set is contracted from $\Delta$ to $\Delta / 8$, the corresponding optimal control protocol converges to $\theta^{\mathrm{w.o.}}$. However, partially sampling uncertainty realizations from a subset of points in $\Delta$ does not account for all possible situations:  it is evident from this plot that using control directly, without any robust optimization can result in a dramatic decrease of fidelity over large regions of the uncertainty set. When the length of the qubit chain increases, the performance of the robust protocol for a fixed uncertainty set $\Delta$ also decreases (Fig. \ref{fig:multi}(c)).

\begin{figure}
    \centering
    \includegraphics[width = 9.5 cm]{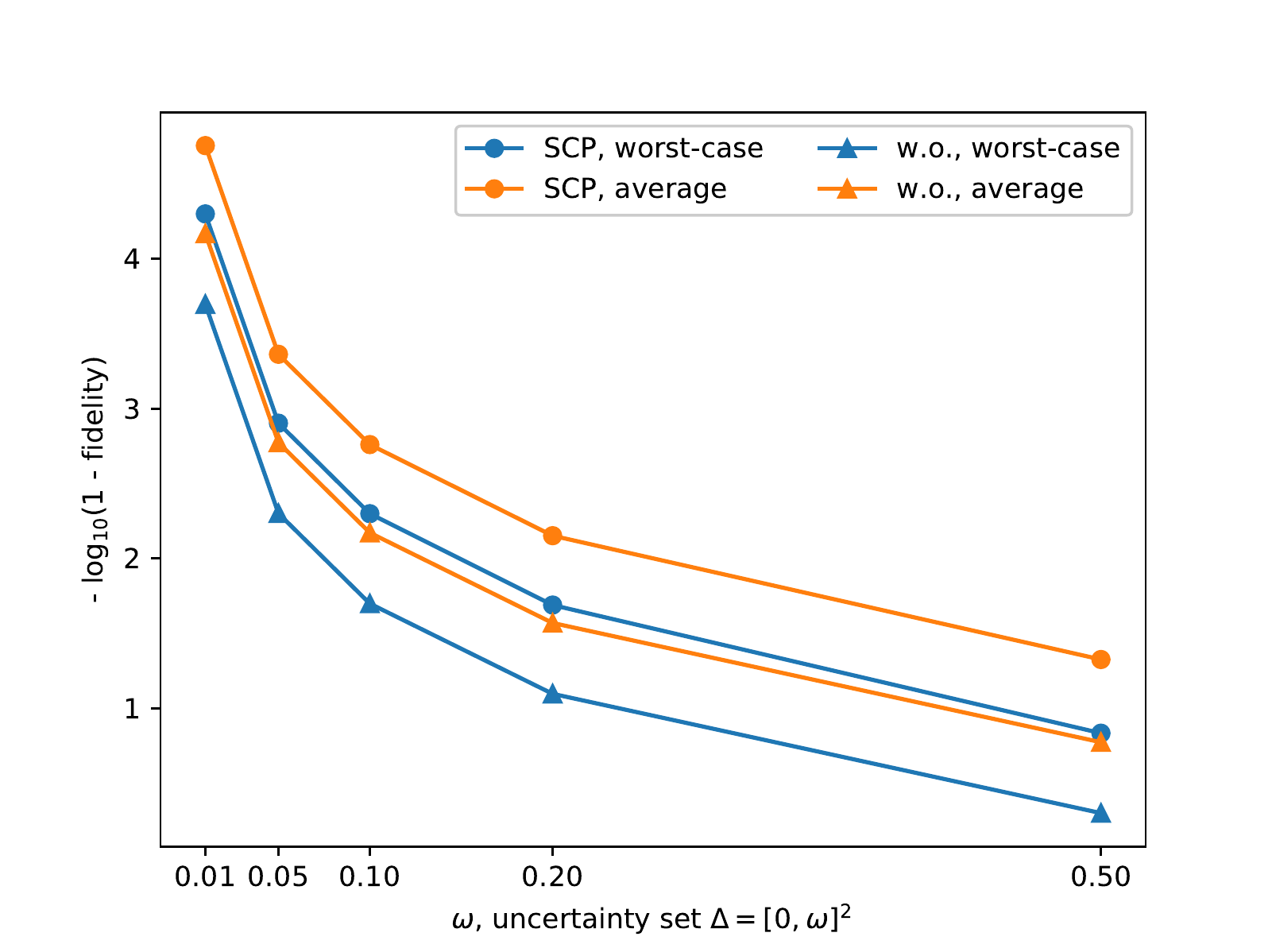}
    \caption{Robust QAOA for seven-qubit chain with uncertainty in initial state preparation. In the figure, circles stand for control angles $\theta^{\mathrm{scp}}$ optimized by SCP. Triangles stand for the control angle $\theta^{\mathrm{w.o.}}$ optimized in the absence of uncertainty until infidelity below 1e-7 and data are calculated by directly turning on uncertainty in the system controlled by $\theta^{\mathrm{w.o.}}$. Both $\theta^{\mathrm{scp}}$ and $\theta^{\mathrm{w.o.}}$ are obtained by starting from the same initial guess.}
    \label{fig:multi-init}
\end{figure}

In addition to studying the robustness to uncertainty in the control Hamiltonians, we also consider the uncertainty when preparing the initial state. To model such uncertainty, we parameterize the initial state as follows:
\begin{equation}
    |\psi_i(\omega_2, \omega_3)\rangle = \sqrt{1 - \omega_2^2 - \omega_3^2} |\bar{1}\rangle + \omega_2 |\bar{2}\rangle + \omega_3 |\bar{3}\rangle.
\end{equation}
Here $\omega_i$ stands for the error contribution from the $i^{th}$ excited state.
Assuming that the control Hamiltonians are uncertainty free, the fidelity of the problem is given by $\mathcal{F}(\theta, \delta) = |\langle \psi_t| U(\theta) | \psi_i(\delta)\rangle|^2$ where $\delta = (\omega_2, \omega_3) \in \Delta$. Robust optimization is then performed to find the optimal control vector as a function of the uncertainty in initial state preparation. To demonstrate the non-trivial contribution of robust optimization, we undertake another optimization from the same initial value of the control variables $\theta^0$, but without turning on uncertainty, until the infidelity attains $10^{-7}$. The resulting control variable are labelled $\theta^{\mathrm{w.o.}}$. Fig.\ref{fig:multi-init} compares the performance of $\theta^{\mathrm{scp}}$ and $\theta^{\mathrm{w.o.}}$. We see that the fidelity is significantly improved by robust optimization for all values of uncertainty $\Delta$. 

\section{Summary}\label{sect:sum}

Quantum variational algorithms, such as
QAOA, can possibly be realized on near term quantum computing machines in a number of scientific applications.  Developing robust realizations of these algorithms that allow
control and mitigation of errors on near term devices to the extent
that a quantum enhancement is asserted, will have significant impact
on both the quantum computing community and its potential users.

In this work we have investigated how robust optimization, in particular SCP, can robustly assist the design of quantum control protocols. Since the performance of near term quantum devices is severely limited by decoherence, the exploitation of the computational capability requires that the quantum circuits should be relatively robust against noise and uncertainties. Using several numerical examples, we demonstrate that the achievable fidelity of QAOA can significantly decrease in the presence of uncertainties in the Hamiltonian. Meanwhile the performance of QAOA may be significantly  enhanced by solving the max-min problem in the framework of robust optimization. Future work will address application of these ideas to mitigate errors deriving from measurements and from coupling of qubits to their environment.

\begin{ack}

{This work was partially supported by a Google Quantum Research Award (Y.D., L.L., B.W.) and by the Quantum Algorithm Teams Program under Grant No. DE-AC02-05CH11231 (L.L. and B.W.). X.M. thanks the Office of international relations, Peking University, Beijing, China for partial funding of an exchange studentship at the University of California, Berkeley.}

\end{ack}

\appendix
\section{Numerical Environments and Details}\label{appendix:detail}
All experiments are performed on an Intel Core 4 Quad CPU at 2.30 GHZ with 8 GB of RAM.  All codes are written in MATLAB R2018b. For each optimization model presented in this paper, we first found a control vector $\theta_0$ that can achieve a fidelity of $0.999$ without uncertainty by GRAPE method. Then we served $\theta_0$ as an initial guess for robust optimization. Since for single-qubit system, the quality of robust solution is largely dependent on the choice of initial guess, we performed each algorithm 10 times starting from 10 different initial values and chose the best solution. 

Throughout the numerical experiments, we utilized the first-order approximation Eq.\eqref{scp:gradapp} for SCP. The parameters for Algorithm \ref{alg:scp} are set as follows:  $\eta_1=0.5$, $\eta_2=0.1$, $\gamma_1=2$, $\gamma_2=0.2$, $\text{tol}_d=$1e-6. $t_{max}$ is set to be 500 for single-qubit system, 2e3 for multi-qubit system (I), and 2e4 for multi-qubit system (II). $\text{tol}_{\sigma}$ is 1e-6 for single-qubit system and 1e-8 for multi-qubit systems. We performed  b-GRAPE  for a fixed number of iterations with $\mathrm{Batchsize}=1$ and momentum parameter $\lambda=0.9$, which are  best parameters adopted by the  original author of b-GRAPE method. The number of iterations is set to be 2e4 for single-qubit system and 5e4 for multi-qubit systems. We implemented the best-response version of a-GRAPE. We performed $15$ rounds of a-GRAPE training for single-qubit system and multi-qubit system (II). While for multi-qubit system (I), the number of rounds is set to be $25$. We would like to note here that for b-GRAPE and a-GRAPE, a larger number of iterations will slightly improve the quality of the solution, but cost more computational resources. The numbers of iterations we chose here are designed to minimize the CPU time while maintaining a high performance.

\section{Comparison of Algorithms}\label{appendix:compare}
 In this section, we will compare SCP, b-GRAPE and a-GRAPE on single-qubit system and multi-qubit systems. The descriptions of these systems have been provided in Sec.\ref{sect:errmitigate}. Here we provide the pseudo-codes of b-GRAPE and a-GRAPE in Algorithm \ref{alg:bgrape} and Algorithm \ref{alg:agrape} respectively for completeness.
 
\begin{algorithm}[!htbp]  
\caption{b-GRAPE method}
\label{alg:bgrape}
\begin{algorithmic}
    \REQUIRE initial guess $\theta_0$,  a positive integer $t_{max}$, momentum parameter $\lambda>0$, batch size $n_{b}$.
    \ENSURE control vector $\theta_{t_{max}}$.
    \STATE Set initial momentum $g_0=0$.
    \FOR{$t = 1,2, \dots,t_{max}$}
    \STATE Choose training samples $\delta_i\in\Delta$, $1\le i\le n_b$ randomly.
    \STATE Update $\theta$ via
    $$
    \begin{aligned}
    g_t &= \frac{1}{n_b}\sum_{i=1}^{n_b}\frac{\partial \mathcal{F}(\theta,\delta_i)}{\partial \theta}+\lambda g_{t-1}, \\
    \theta_t &= \theta_{t-1}+\beta g_t,
    \end{aligned}
    $$ 
    
    where  $\beta$ is the learning rate.
    \ENDFOR
\end{algorithmic}
\end{algorithm}
 
\begin{algorithm}[!htbp]  
\caption{a-GRAPE method}
\label{alg:agrape}
\begin{algorithmic}
    \REQUIRE initial guess $\theta_0$,  a positive integer $t_{max}$, a set memory size $s=10$.
    \ENSURE control vector $\theta_{t_{max}}$.
    \STATE Set initial uncertainty set $B_0=\{0\}$. 
    \FOR{$t = 1,2, \dots,t_{max}$}
    \STATE (1) Find an approximate optimal control $\theta_t$ over the uncertainty set $B_{t-1}$ by GRAPE method
    $$
    \theta_t \approx \argmax_{\theta} \sum_{\delta\in B_{t-1}} \mathcal{F}(\theta,\delta).
    $$
    \STATE (2) Generate an approximate adversarial noise instance by
    $$
    \delta_t \approx\argmin_{\delta\in\Delta} \mathcal{F}(\theta_t,\delta).
    $$ 
    \STATE (3) Update uncertainty set $B$ by following rules:
    \IF{$|B_{t-1}|< s$}
       \STATE $$
        B_t= B_{t-1}\cup\{\delta_t\},
        $$
    \ELSE
       \STATE $$
        B_t=\{\delta_{t-s+1},\dots,\delta_{t}\}.
        $$
    \ENDIF
    \ENDFOR
\end{algorithmic}
\end{algorithm}
The numerical results are displayed in Tab.\ref{tab:numeric}. We see the advantages of our proposed method in both accuracy and CPU time. The control protocols obtained by SCP have the lowest worst-case infidelity in most cases and are close to the best one in the other situations. As to the speed, a-GRAPE and b-GRAPE are 2-5 times slower than SCP in general. Besides, although we design the optimization objective function Eq.\eqref{eq:control} in order to maximize the worst-case fidelity, SCP can nonetheless attain a satisfying average fidelity in practice. Thus, it is evident that finding a robust QAOA control via SCP is efficient and accurate. We ascribe these advantages to the trust-region scheme and sample tactic utilized in SCP. Adaptively adjusting the size of trust region on the fly and rejecting unsatisfying updates guarantee the efficiency of SCP. Moreover, approximating the uncertainty via finite samples provides a simple but accurate way to deal with optimization over the bounded uncertainty set.

\begin{table*}[]
\centering
\small
\caption{Numerical results for single-qubit and multi-qubit systems in Sec.\ref{sect:errmitigate}. Here w-c and avg stand for worst-case and average infidelity respectively. Time is the CPU time in second taken by each optimization. The best results (in terms of accuracy or timing) are in boldface.\label{tab:numeric}} 
\vspace{0.3cm}

\subtable[Numerical results for single-qubit system in Sec.\ref{sect:numeric:sing}.\label{tab:sing}]{  
       \begin{tabular}{|c|c|rrr|rrr|rrr|}
\hline
\multicolumn{2}{|c|}{\multirow{2}{*}{\tabincell{c}{QAOA depth $p$\\Uncertainty set $\Delta$}}} & \multicolumn{3}{c|}{SCP} & \multicolumn{3}{c|}{b-GRAPE} & \multicolumn{3}{c|}{a-GRAPE} \\ \cline{3-11}
\multicolumn{2}{|c|}{} & \multicolumn{1}{c}{w-c} & \multicolumn{1}{c}{avg} & \multicolumn{1}{c|}{time(s)} & \multicolumn{1}{c}{w-c} & \multicolumn{1}{c}{avg} & \multicolumn{1}{c|}{time(s)} & \multicolumn{1}{c}{w-c} & \multicolumn{1}{c}{avg} & \multicolumn{1}{c|}{time(s)} \\ \hline
\multirow{3}{*}{$p=5$} & $\Delta_1$ & \textbf{8.83e-04} & \textbf{5.91e-04} & \textbf{2.06e+02} & 1.72e-02 & 3.13e-03 & 3.71e+02 & 9.00e-03 & 3.37e-03 & 4.52e+02 \\ 
 & $\Delta_2$ & \textbf{1.26e-03} & \textbf{4.48e-04} & \textbf{1.65e+02} & 1.12e-02 & 3.11e-03 & 3.63e+02 & 5.80e-03 & 1.31e-03 & 4.45e+02 \\ 
 & $\Delta_3$ & \textbf{2.90e-02} & 2.19e-02 & \textbf{1.12e+02} & 7.32e-02 & \textbf{1.33e-02} & 3.63e+02 & 3.84e-02 & 2.25e-02 & 4.40e+02 \\ \hline
\multirow{3}{*}{$p=6$} & $\Delta_1$ & \textbf{1.87e-04} & \textbf{7.86e-05} & \textbf{2.44e+02} & 8.94e-03 & 2.13e-03 & 4.36e+02 & 3.82e-03 & 1.97e-03 & 5.38e+02 \\ 
 & $\Delta_2$ & \textbf{1.54e-03} & \textbf{8.20e-04} & \textbf{2.40e+02} & 3.01e-02 & 5.16e-03 & 4.35e+02 & 6.66e-03 & 2.46e-03 & 5.33e+02 \\ 
 & $\Delta_3$ & \textbf{1.30e-02} & \textbf{7.12e-03} & \textbf{2.14e+02} & 6.93e-02 & 1.86e-02 & 4.35e+02 & 3.34e-02 & 1.68e-02 & 5.33e+02 \\ \hline
\multirow{3}{*}{$p=7$} & $\Delta_1$ & \textbf{7.83e-05} & \textbf{3.38e-05} & \textbf{2.86e+02} & 1.06e-02 & 2.60e-03 & 5.01e+02 & 2.88e-03 & 9.15e-04 & 6.16e+02 \\ 
 & $\Delta_2$ & \textbf{5.20e-04} & \textbf{1.09e-04} & \textbf{2.68e+02} & 1.65e-02 & 2.57e-03 & 5.42e+02 & 8.87e-03 & 2.05e-03 & 6.56e+02 \\ 
 & $\Delta_3$ & \textbf{4.14e-03} & \textbf{1.30e-03} & \textbf{2.78e+02} & 5.03e-02 & 1.11e-02 & 4.95e+02 & 1.86e-02 & 7.15e-03 & 6.06e+02 \\ \hline
\multirow{3}{*}{$p=8$} & $\Delta_1$ & \textbf{2.49e-05} & \textbf{8.37e-06} & \textbf{3.10e+02} & 1.49e-02 & 2.44e-03 & 5.66e+02 & 2.43e-03 & 7.65e-04 & 6.97e+02 \\ 
 & $\Delta_2$ & \textbf{7.88e-05} & \textbf{2.81e-05} & \textbf{3.04e+02} & 1.55e-02 & 2.11e-03 & 5.71e+02 & 4.52e-03 & 2.42e-03 & 7.04e+02 \\ 
 & $\Delta_3$ & \textbf{6.88e-04} & \textbf{3.45e-04} & \textbf{3.30e+02} & 3.66e-02 & 9.27e-03 & 5.76e+02 & 1.33e-02 & 6.89e-03 & 7.08e+02 \\ \hline
\multirow{3}{*}{$p=9$} & $\Delta_1$ & \textbf{2.12e-06} & \textbf{9.08e-07} & \textbf{3.48e+02} & 3.70e-03 & 8.41e-04 & 6.43e+02 & 1.05e-03 & 2.99e-04 & 7.84e+02 \\ 
 & $\Delta_2$ & \textbf{1.92e-05} & \textbf{6.43e-06} & \textbf{3.79e+02} & 1.29e-02 & 1.91e-03 & 6.40e+02 & 7.50e-03 & 1.83e-03 & 7.85e+02 \\ 
 & $\Delta_3$ & \textbf{1.69e-04} & \textbf{9.04e-05} & \textbf{3.63e+02} & 2.95e-02 & 6.68e-03 & 6.47e+02 & 1.23e-02 & 5.19e-03 & 7.97e+02 \\ \hline
\end{tabular}
}  
\vspace{1cm}

\qquad  
\subtable[Numerical results for multi-qubit system (I) in Sec.\ref{sect:numeric:manipulation}.\label{tab:manipulation}]{          
 \resizebox{18cm}{!}{\begin{tabular}{|c|l|rrr|rrr|rrr|}
\hline
\multicolumn{2}{|c|}{\multirow{2}{*}{\tabincell{c}{System size $N$\\Uncertainty set $\Delta$}}} & \multicolumn{3}{c|}{SCP} & \multicolumn{3}{c|}{b-GRAPE} & \multicolumn{3}{c|}{a-GRAPE} \\ \cline{3-11}
\multicolumn{2}{|c|}{} & \multicolumn{1}{c}{w-c} & \multicolumn{1}{c}{avg} & \multicolumn{1}{c|}{time(s)} & \multicolumn{1}{c}{w-c} & \multicolumn{1}{c}{avg} & \multicolumn{1}{c|}{time(s)} & \multicolumn{1}{c}{w-c} & \multicolumn{1}{c}{avg} & \multicolumn{1}{c|}{time(s)} \\ \hline
\multirow{4}{*}{$N=3$} & $\Delta_1=[-0.01,0.01]^2$ & \textbf{6.40e-04} & \textbf{2.06e-04} & \textbf{2.46e+01} & 4.92e-03 & 8.83e-04 & 5.08e+01 & 8.73e-04 & 2.82e-04 & 1.39e+02 \\ 
 & $\Delta_2=[-0.02,0.02]^2$ & \textbf{1.31e-03} & \textbf{4.72e-04} & \textbf{4.69e+01} & 9.13e-03 & 2.01e-03 & 4.70e+01 & 2.43e-03 & 1.01e-03 & 1.39e+02 \\ 
 & $\Delta_3=[-0.05,0.05]^2$ & \textbf{1.68e-03} & \textbf{8.98e-04} & 1.04e+02 & 1.62e-02 & 3.10e-03 & \textbf{4.67e+01} & 5.05e-03 & 1.59e-03 & 1.37e+02 \\ 
 & $\Delta_4=[-0.1,0.1]^2$ & \textbf{3.54e-03} & \textbf{2.18e-03} & 8.35e+01 & 3.66e-02 & 4.46e-03 & \textbf{4.67e+01} & 8.56e-03 & 4.97e-03 & 1.37e+02 \\ \hline
\multirow{4}{*}{$N=4$} & $\Delta_1=[-0.01,0.01]^2$ & \textbf{2.26e-03} & 7.20e-04 & \textbf{3.20e+01} & 5.99e-03 & 1.51e-03 & 1.09e+02 & 2.33e-03 & \textbf{6.11e-04} & 2.98e+02 \\ 
 & $\Delta_2=[-0.02,0.02]^2$ & \textbf{3.07e-03} & \textbf{1.09e-03} & 1.18e+02 & 8.46e-03 & 2.15e-03 & \textbf{1.09e+02} & 5.82e-03 & 1.84e-03 & 2.99e+02 \\ 
 & $\Delta_3=[-0.05,0.05]^2$ & \textbf{6.19e-03} & \textbf{2.18e-03} & 2.00e+02 & 2.50e-02 & 6.01e-03 & \textbf{1.08e+02} & 1.50e-02 & 5.38e-03 & 3.00e+02 \\ 
 & $\Delta_4=[-0.1,0.1]^2$ & \textbf{1.18e-02} & \textbf{6.78e-03} & \textbf{8.89e+01} & 7.02e-02 & 1.48e-02 & 1.08e+02 & 2.79e-02 & 1.13e-02 & 3.00e+02 \\ \hline
\multirow{4}{*}{$N=5$} & $\Delta_1=[-0.01,0.01]^2$ & \textbf{5.63e-03} & 1.97e-03 & \textbf{1.80e+02} & 1.21e-02 & 2.80e-03 & 3.79e+02 & 5.73e-03 & \textbf{1.78e-03} & 1.40e+03 \\ 
 & $\Delta_2=[-0.02,0.02]^2$ & \textbf{1.42e-02} & \textbf{5.01e-03} & \textbf{2.75e+02} & 2.32e-02 & 6.68e-03 & 3.79e+02 & 1.67e-02 & 5.66e-03 & 1.41e+03 \\ 
 & $\Delta_3=[-0.05,0.05]^2$ & \textbf{3.26e-02} & \textbf{1.23e-02} & \textbf{3.50e+02} & 9.02e-02 & 2.22e-02 & 3.83e+02 & 5.77e-02 & 2.07e-02 & 1.40e+03 \\ 
 & $\Delta_4=[-0.1,0.1]^2$ & 1.50e-01 & 9.76e-02 & \textbf{2.28e+02} & 3.40e-01 & \textbf{7.49e-02} & 3.83e+02 & \textbf{1.46e-01} & 7.72e-02 & 1.39e+03 \\ \hline
\multirow{4}{*}{$N=6$} & $\Delta_1=[-0.01,0.01]^2$ & 1.25e-02 & 3.63e-03 & \textbf{4.01e+02} & 2.32e-02 & 5.05e-03 & 2.79e+03 & \textbf{1.19e-02} & \textbf{3.55e-03} & 9.79e+03 \\ 
 & $\Delta_2=[-0.02,0.02]^2$ & \textbf{2.07e-02} & \textbf{7.38e-03} & \textbf{1.10e+03} & 5.54e-02 & 1.26e-02 & 2.80e+03 & 2.63e-02 & 8.69e-03 & 9.75e+03 \\ 
 & $\Delta_3=[-0.05,0.05]^2$ & 8.39e-02 & 3.38e-02 & \textbf{1.27e+03} & 1.46e-01 & 4.46e-02 & 2.80e+03 & \textbf{7.90e-02} & \textbf{2.94e-02} & 9.77e+03 \\ 
 & $\Delta_4=[-0.1,0.1]^2$ & \textbf{2.35e-01} & 1.46e-01 & \textbf{1.24e+03} & 4.11e-01 & \textbf{1.06e-01} & 2.78e+03 & 2.69e-01 & 1.26e-01 & 9.77e+03 \\ \hline
\end{tabular}
}}
\vspace{1cm}

\qquad 
\subtable[Numerical results for multi-qubit system (II) in Sec.\ref{sect:numeric:multi} in which uncertainty involves in the initial state.\label{tab:init}]{
\begin{tabular}{|l|rrr|rrr|rrr|}
\hline
\multirow{2}{*}{Uncertainty set} & \multicolumn{3}{c|}{SCP} & \multicolumn{3}{c|}{b-GRAPE} & \multicolumn{3}{c|}{a-GRAPE} \\ \cline{2-10}
 & \multicolumn{1}{c}{w-c} & \multicolumn{1}{c}{avg} & \multicolumn{1}{c|}{time(s)} & \multicolumn{1}{c}{w-c} & \multicolumn{1}{c}{avg} & \multicolumn{1}{c|}{time(s)} & \multicolumn{1}{c}{w-c} & \multicolumn{1}{c}{avg} & \multicolumn{1}{c|}{time(s)} \\ \hline
$\Delta_1=[0,0.01]^2$ & \textbf{5.00e-05} & 1.74e-05 & \textbf{1.01e+02} & 1.05e-04 & 5.47e-05 & 6.72e+02 & 5.02e-05 & \textbf{1.74e-05} & 7.27e+02 \\

$\Delta_2=[0,0.05]^2$ & \textbf{1.25e-03} & \textbf{4.34e-04} & \textbf{1.10e+02} & 1.36e-03 & 4.73e-04 & 6.79e+02 & 1.26e-03 & 4.34e-04 & 7.32e+02 \\

$\Delta_3=[0,0.1]^2$ & \textbf{5.03e-03} & \textbf{1.74e-03} & \textbf{1.06e+02} & 5.52e-03 & 1.79e-03 & 6.76e+02 & 5.03e-03 & 1.74e-03 & 7.45e+02 \\

$\Delta_4=[0,0.2]^2$ & \textbf{2.04e-02} & 7.01e-03 & \textbf{1.01e+02} & 2.12e-02 & 7.05e-03 & 6.69e+02 & 2.05e-02 & \textbf{7.01e-03} & 7.38e+02 \\

$\Delta_5=[0,0.5]^2$ & \textbf{1.46e-01} & 4.72e-02 & \textbf{6.76e+01} & 1.62e-01 & \textbf{4.68e-02} & 6.73e+02 & 1.48e-01 & 4.73e-02 & 7.27e+02 \\ \hline
\end{tabular}
}  
\end{table*}

\end{document}